# Neural mass modeling of slow-fast dynamics of seizure initiation and abortion

Elif Köksal Ersöz[1], Julien Modolo[1], Fabrice Bartolomei[2, 3], Fabrice Wendling[1*]

1 University of Rennes, Inserm-U1099, LTSI, Rennes, France

2 Aix Marseille University, Inserm, INS, Institut de Neurosciences des Systèmes, Marseille, France

3 APHM, Timone Hospital, Clinical Neurophysiology, Marseille, France

* Corresponding author

E-mail: fabrice.wendling@inserm.fr

## Abstract

Epilepsy is a dynamic and complex neurological disease affecting about 1% of the worldwide population, among which 30% of the patients are drug-resistant. Epilepsy is characterized by recurrent episodes of paroxysmal neural discharges (the so-called seizures), which manifest themselves through a large-amplitude rhythmic activity observed in depth-EEG recordings, in particular in local field potentials (LFPs). The signature characterizing the transition to seizures involves complex oscillatory patterns, which could serve as a marker to prevent seizure initiation by triggering appropriate therapeutic neurostimulation methods. To investigate such protocols, neurophysiological lumped-parameter models at the mesoscopic scale, namely neural mass models, are powerful tools that not only mimic the LFP signals but also give insights on the neural mechanisms related to different stages of seizures. Here, we analyze the multiple time-scale dynamics of a neural mass model and explain the underlying structure of the complex oscillations observed before seizure initiation. We investigate population-specific effects of the stimulation and the dependence of stimulation parameters on synaptic timescales. In particular, we show that intermediate stimulation frequencies (>20 Hz) can abort seizures if the timescale difference is pronounced. Those results have the potential in the design of therapeutic brain stimulation protocols based on the neurophysiological properties of tissue.

## Author summary

Epilepsy is a complex disease affecting 1% of the worldwide population of which 30% of the patients are drug-resistant and seeking for alternative therapeutic methods, such as neurostimulation. Epileptic seizures are hallmarked by preceding pre-ictal phases which are a possible window of opportunity to trigger electrical stimulation with the objective to prevent seizure initiation. Biophysiological models are an appropriate framework to understand underlying dynamics and transitions between different epileptogenic phases. In this study, we consider a typical pre-ictal regime with complex bursting-type oscillations, which can be accurately reproduced by a neural mass model. By analyzing the multiple time-scaled structure of the model, we identify the key role of the subpopulations of GABAergic interneurons. We show that appropriate brain stimulation targeting GABAergic interneurons is able to abort pre-ictal bursting, thus preventing seizures to develop.

## Introduction

Epilepsy is a severe, multi-causal chronic disease defined by the recurrence of unpredictable seizures that severely affect patients' quality of life. In 30% of patients, antiepileptic drugs [1] remain inefficient to control the occurrence of seizures. In most cases, drug-resistant epilepsies are 'focal', as characterized by an epileptogenic zone (EZ) that is relatively circumscribed in one of the two cerebral hemispheres. There is a large body of evidence supporting that the balance between excitatory and inhibitory processes is modified in the EZ [2] due to multiple, not mutually exclusive, pathological mechanisms resulting from changes occurring at the cellular level (e.g. hyperexcitability caused by potassium and chloride dysregulation, review in [3]), up to the network level (e.g. hyperexcitability caused by altered glutamatergic or GABAergic synaptic transmission, review in [4]). Unfortunately, surgical treatment can only be offered to 15-20% drug-resistant patients [5] in whom the benefit-to-deficit ratio is favorable. Therefore, alternative therapeutic procedures aimed at reducing seizures' frequency are urgently needed.

Among these procedures, direct electrical stimulation of the brain is an increasingly popular technique of treating epilepsy, as evidenced by both animal and human studies [6].

Stimulation targets have included deep brain structures such as thalamic nuclei, hippocampus or cortical targets [7]. It has been acknowledged for decades that stimulation of the cortex during routine brain mapping procedures may induce epileptiform discharges or seizures, but more recently pulse trains have demonstrated their potential in aborting abnormal epileptiform activity [8]. Direct stimulation has been shown to be effective in suppressing epileptic activity, however with inconsistent results among patients. Furthermore, brain stimulation in drug-refractory patients is recognized to be still largely empirical [9]. A rational definition of stimulation protocols is indeed still missing, as evidenced in randomized controlled trials [10].

In this context, the specific objective of the present study is to exploit neuro-inspired models to design neurostimulation protocols aiming at aborting seizures at their onset. More specifically, we investigate a well-defined pattern of interictal-to-ictal transition characterized by the occurrence of pre-ictal rhythmic large amplitude spikes followed by a fast onset activity, as observed in stereo-EEG recordings (SEEG, intracerebral electrodes). Although not the unique one, this commonly encountered pattern has long been considered as a hallmark of the EZ, especially in mesial temporal lobe epilepsy [11–13]. More particularly, we focus on pre-ictal bursting characterized by active episodes (fast epileptic spikes), repeated (quasi-) periodically and separated by quiescent (slow-wave and/or silent) phases. First, we accurately reproduce human electrophysiological patterns in a neural mass model featuring glutamatergic pyramidal neurons as well as two types of GABAergic interneurons (somatostatin-positive or SOM+, and parvalbumin-positive or PV+). After integrating neurostimulation effects in the model as a parametrizable exogenous membrane perturbation of the main cells and interneurons, we analyze the slow-fast nature of this nonlinear dynamical system in the bursting regime by using numerical bifurcation analysis and geometric singular perturbation theory (GSPT) [14,15]. Following this approach, the mechanisms leading to the pre-ictal bursting are determined, and the perturbation effects are explained geometrically. Identification of the model structures to be targeted for bursting abortion highlight the key role of SOM+ interneurons in suppressing pre-ictal epileptic activity. Overall, the results of this study elucidate the nature of pre-ictal spike bursting and provide key information to design optimal direct stimulation protocols targeting this specific epileptiform pattern.

## Model and Methods

### Model

We consider the neural mass model presented in [16] which includes three interacting neuronal subpopulations: pyramidal neurons and inhibitory interneurons (SOM+ and PV+, also called "dendrite-projecting slow" and "soma-projecting fast" interneurons, respectively). The average postsynaptic potential of each subpopulation is determined by two functions: 1) a 'pulse-to-wave' function, $S(v) = 5/(1 + \exp(0.56(6 - v)))$, transforming the incoming postsynaptic potentials into a firing rate; and 2) the input firing rate is converted into the mean post-synaptic potential of the corresponding subpopulation by a linear transformation, that is $h(t) = Wt/\tau_w \exp(-t/\tau_w)$, where $W$ represents the average synaptic gain and $\tau_w$ is the average synaptic time constant. The system reads:

$$
\begin{aligned}
\ddot{y}_0 &= \frac{A}{\tau_a} S(y_1 - y_2 - y_3) - \frac{2}{\tau_a}\dot{y}_0 - \frac{1}{\tau_a^2}y_0, \\
\ddot{y}_1 &= \frac{A}{\tau_a}\{p(t) + C_2 S(C_1 y_0)\} - \frac{2}{\tau_a}\dot{y}_1 - \frac{1}{\tau_a^2}y_1, \\
\ddot{y}_2 &= \frac{B}{\tau_b} C_4 S(C_3 y_0) - \frac{2}{\tau_b}\dot{y}_2 - \frac{1}{\tau_b^2}y_2, \\
\ddot{y}_3 &= \frac{G}{\tau_g} C_7 S(C_5 y_0 - C_6 y_4) - \frac{2}{\tau_g}\dot{y}_3 - \frac{1}{\tau_b^2}y_3 \\
\ddot{y}_4 &= \frac{B}{\tau_b} S(C_3 y_0) - \frac{2}{\tau_b}\dot{y}_4 - \frac{1}{\tau_b^2}y_4.
\end{aligned}
\tag{1}
$$

Variables $y_i$ stand for the post-synaptic potentials (PSPs) generated at the level of pyramidal cells ($y_0$), excitatory inputs on pyramidal cells ($y_1$), SOM+ interneurons ($y_2$), PV+ interneurons ($y_3$), and inhibitory inputs on PV+ interneurons ($y_4$). Parameters $A, B, G$ are the synaptic gains, $\tau_a, \tau_b, \tau_g$ are the synaptic time constants, connectivity constants $C_i$s represent the average number of synaptic contacts, and $p(t)$ is the external (noisy) cortical input ($p(t) = p + \xi$, where $p$ is the mean of the external input, and $\xi$ is a random variable following a normal distribution with zero mean and standard deviation $\sigma$). Table 1 presents the parameter values used in this manuscript unless otherwise stated. The main difference between the parameter sets of [16] and Table 1 is the connectivity strengths of the circuit involving PV+ interneurons. Note that $\tau_a$ is 3.3 times and $\tau_b$ is 16.6 times greater than $\tau_g$. These differences would introduce multiple time-scale dynamics in the system. Below, we recall primaries of slow-fast analysis before expressing (1) in slow-fast formulation.

**Table 1 Parameter values during simulated background activity**

| $A$ (mV) | $B$ (mV) | $G$ (mV) | $p$ (Hz) | $C_1$ | $C_2$ | $C_3$ | $C_4$ | $C_5$ | $C_6$ | $C_7$ | $\tau_a$(s) | $\tau_b$ (s) | $\tau_g$ (s) |
|---|---|---|---|---|---|---|---|---|---|---|---|---|---|
| 5 | 40 | 35 | 90 | 135 | 108 | 35 | 25 | 450 | 121 | 121 | 0.01 | 0.05 | 0.003 |

**Primaries of slow-fast analysis**

A slow-fast system in the general slow form reads,

$$\epsilon \dot{x} = f(x, z, \epsilon),$$
$$\dot{z} = g(x, z, \epsilon),$$

with fast variables $x$ and slow variables $z$ of arbitrary dimensions, time scale parameter $0 < \epsilon \ll 1$, and dot represents derivation with respect to time $t$. The dynamics of a slow-fast system can be divided into *fast* and *slow* epochs. Each of these epochs can be investigated with the slow-fast analysis in a hybrid manner and then can be concatenated, so that one can understand the underlying structure giving sharp transitions (excitable responses to external inputs) and complex oscillatory patterns (spiking, bursting and subthreshold oscillations) [17].

An important geometrical object for both the slow and the fast dynamics is the *critical manifold* $\mathcal{C}^0$, defined as the nullcline of the fast variable $\mathcal{C}^0 = \{(x,z)|f(x,z,0)=0\}$, eventually obtained by setting $\epsilon = 0$. For the differential-algebraic system defined for $\epsilon = 0$, the so-called *reduced system (slow subsystem)* approximates the slow dynamics of the original system. The critical manifold $\mathcal{C}^0$ both defines the phase space of the reduced system and equilibrium points of the layer problem expressed in the fast time-scale, that is

$$x' = f(x,z,0),$$
$$z' = 0,$$

where ($'$) denotes derivative with respect to $\tau = t/\epsilon$. The stability of the layer problem determines the characteristics of the critical manifold. The critical manifold $\mathcal{C}^0$ is normally hyperbolic along the set for which $\det(f_x(x,z,0)) \neq 0$, which can be attracting, repelling or saddle type. The Fenichel theory [14] guarantees that these normally hyperbolic points of the critical manifold perturb smoothly in $\epsilon$ and give *slow manifolds* ($\mathcal{C}^\epsilon$) of the original system for small enough $\epsilon > 0$. If $\mathcal{C}$ is folded, attracting and repelling branches of $\mathcal{C}^0$ meet along the fold set $\mathcal{F} = \{\det(f_x(x_{fold}, z_{fold}, 0)) = 0\}$, where normal hyperbolicity is lost. Extension of the classical Fenichel theory to non-hyperbolic sets provides a tool to investigate the slow dynamics near $\mathcal{F}$, and one can expect *canard solutions* in the neighborhood of such sets [18].

**Slow-fast formulation of the model**

One can notice that the variable $y_4$ in (1) is equivalent to $y_2$, thus the dimension of (1) can be reduced by multiplying the post-synaptic potential variables with $C_i$s before the 'pulse-to-wave' conversion. Further, by applying the variable conversion,

$$\left(\frac{y_0}{\tau_a}, \frac{y_1}{\tau_a}, \frac{y_2}{\tau_b}, \frac{y_3}{\tau_g}, y_5, y_6, y_7, y_8\right) \rightarrow (v_0, v_1, v_2, v_3, v_4, y_5, y_6, y_7, y_8),$$

system (1) can be written as:

$$
\begin{aligned}
\tau_g \dot{v}_3 &= y_8, \\
\tau_g \dot{y}_8 &= G\,S(C_5 \tau_a v_0 - C_6 \tau_b v_2) - v_3 - 2y_8, \\
\tau_a \dot{v}_0 &= y_5, \\
\tau_a \dot{y}_5 &= A\,S\big(A\tau_a p + C_2 \tau_a v_1 - C_4 \tau_b v_2 - C_7 \tau_g v_3\big) - v_0 - 2y_5, \\
\tau_a \dot{v}_1 &= y_6, \\
\tau_a \dot{y}_6 &= A\,S(C_1 \tau_a v_0) - v_1 - 2y_6, \\
\tau_b \dot{v}_2 &= y_7, \\
\tau_b \dot{y}_7 &= B\,S(C_3 \tau_a v_0) - v_2 - 2y_7.
\end{aligned}
\tag{2}
$$

Intuitively, system (1), hence system (2), are multiple-time-scale systems which can result in complex epileptogenic patterns for appropriate choices of parameters. Thus, understanding the multiple-time-scale structure of (2) is indispensable for designing brain stimulation protocols aiming at aborting the aforementioned oscillatory patterns. In order to proceed a slow-fast analysis of (2) and use the standard methods of GSPT, we define two parameters, $\delta = \tau_g/\tau_a$ and $\varepsilon = \tau_a/\tau_b$. We further assume that $\varepsilon$ and $\delta$ are independent of the synaptic time constants $(\tau_a, \tau_b, \tau_g)$, similar to the approach followed in [19]. After normalizing time with respect to $\tau_g$, as $\tilde{t} = t/\tau_g$, system (2) is expressed in an explicit slow-fast formulation:

$$\frac{dv_3}{d\tilde{t}} = y_8 := F_3(y_8),$$

$$\frac{dy_8}{d\tilde{t}} = G\,S(C_5\tau_a v_0 - C_6\tau_b v_2) - v_3 - 2y_8 := F_8(v_0, v_2, v_3, y_8),$$

$$\frac{dv_0}{d\tilde{t}} = \delta y_5 := \delta F_0(y_5),$$

$$\frac{dy_5}{d\tilde{t}} = \delta\big(A\,S(A\tau_a p + C_2\tau_a v_1 - C_4\tau_b v_2 - C_7\tau_g v_3) - v_0 - 2y_5\big) := F_5(v_0, v_1, v_2, v_3, y_5),$$

$$\frac{dv_1}{d\tilde{t}} = \delta y_6 := \delta F_1(y_6),$$

$$\frac{dy_6}{d\tilde{t}} = \delta(A\,S(C_1\tau_a v_0) - v_1 - 2y_6) := \delta F_6(v_0, v_1, y_6),$$

$$\frac{dv_2}{d\tilde{t}} = \delta\varepsilon y_7 := \delta\varepsilon F_2(y_7),$$

$$\frac{dy_7}{d\tilde{t}} = \delta\varepsilon(B\,S(C_3\tau_a v_0) - v_2 - 2y_7) := \delta\varepsilon F_7(v_0, v_2, y_7). \tag{3}$$

System (3) is a three-time-scale system for small enough values of $(\delta, \varepsilon)$ [27–31], with $(v_3, y_8)$ being fast variables, $(v_0, y_5, v_1, y_6)$ slow variables, and $(v_2, y_7)$ superslow variables. System (3) is written using the (fast) time $\tilde{t}$, and called the *fast system*. We follow [20,21] to analyze the three time scaled slow-fast structure of (3). Defining $\tilde{t}_s = \delta\tilde{t}$ gives the *slow system*:

$$\delta\frac{dv_3}{d\tilde{t}_s} = F_3(y_8)$$

$$\delta\frac{dy_8}{d\tilde{t}_s} = F_8(v_0, v_2, v_3, y_8),$$

$$\frac{dv_0}{d\tilde{t}_s} = F_0(y_5),$$

$$\frac{dy_5}{d\tilde{t}_s} = F_5(v_0, v_1, v_2, v_3, y_5),$$

$$\frac{dv_1}{d\tilde{t}_s} = F_1(y_6),$$

$$\frac{dy_6}{d\tilde{t}_s} = F_6(v_0, v_1, y_6),$$

$$\frac{dv_2}{d\tilde{t}_s} = \varepsilon F_2(y_7),$$

$$\frac{dy_7}{d\tilde{t}_s} = \varepsilon F_7(v_0, v_2, y_7). \tag{4}$$

where $F_i$s are as defined for (3), with $i$ representing the system variables' indices on the left-hand side. **S1** Fig(a) presents the bifurcation diagram of the $(v_3, y_8, v_0, y_5, v_1, v_6)$-subsystem of (4) for $\varepsilon = 0$, where $v_2$ acts as a parameter, and a periodic bursting orbit of (3) for $\varepsilon = 0.01$. We see that the orbit agrees with the bifurcation diagram when $\varepsilon$ is decreased. Details of the bursting behavior are explained in Sec. *Bursting analysis.*

Defining $\tilde{t}_{ss} = \varepsilon\tilde{t}_s = \varepsilon\delta\tilde{t}$ gives *the superslow system,*

$$\varepsilon\delta\frac{dv_3}{d\tilde{t}_{ss}} = F_3(y_8),$$
$$\varepsilon\delta\frac{dy_8}{d\tilde{t}_{ss}} = F_8(v_0, v_2, v_3, y_8),$$
$$\varepsilon\frac{dv_0}{d\tilde{t}_{ss}} = F_0(y_5),$$
$$\varepsilon\frac{dy_5}{d\tilde{t}_{ss}} = F_5(v_0, v_1, v_2, v_3, y_5),$$
$$\varepsilon\frac{dv_1}{d\tilde{t}_{ss}} = F_1(y_6),$$
$$\varepsilon\frac{dy_6}{d\tilde{t}_{ss}} = F_6(v_0, v_1, y_6),$$
$$\frac{dv_2}{d\tilde{t}_{ss}} = F_2(y_7),$$
$$\frac{dy_7}{d\tilde{t}_{ss}} = F_7(v_0, v_1, y_7).$$

$$(5)$$

Systems (3), (4) and (5) are equivalent if $\varepsilon \neq 0$ and $\delta \neq 0$, but they give nonequivalent dynamics in the singular limits $\varepsilon \to 0$ and/or $\delta \to 0$. The limit $\delta \to 0$ in the fast system (3) eliminates the slow and superslow dynamics and yields the *fast layer problem,*

$$\frac{dv_3}{d\tilde{t}} = F_3(y_8),$$
$$\frac{dy_8}{d\tilde{t}} = F_8(v_0, v_2, v_3, y_8),$$

$$(6)$$

which describes the dynamics of the fast variables $(v_3, y_8)$ for fixed values of $(v_0, v_2)$, $(v_0^0, v_2^0)$ for instance. The critical manifold is defined by the four-dimensional set of equilibria of the fast layer problem (6), which reads,

$$S^0 = \{(v_3, y_8, v_0, v_2) \mid F_3(y_8) = 0 \cap F_8(v_0, v_2, v_3, y_8) = 0\},$$

and $S^0$ is eventually in the $(y_8 = 0)$-space. The stability of $S^0$ is determined by deriving the Jacobian of $S^0$ with respect to the fast variables, that is,

$$\text{Jac}\left(S^0_{v_3, y_8}\right) = \begin{bmatrix} 0 & 1 \\ -1 & -2 \end{bmatrix}.$$

Since $\det\left(Jac\left(S^0_{v_3, y_8}\right)\right) \neq 0$, and the eigenvalues are $\lambda_{1,2} = -1$, the $S^0$ is normally hyperbolic and stable. Hence, $S^0$ is perturbed to local invariant slow manifolds for sufficiently small $\delta > 0$.

Another singular limit is obtained by letting $\delta \to 0$ in the slow system (4) gives the algebraic-differential *slow reduced problem*,

$$
\begin{aligned}
0 &= F_3(y_8), \\
0 &= F_8(v_0, v_2, v_3, y_8), \\
\frac{dv_0}{d\tilde{t}_s} &= F_0(y_5), \\
\frac{dy_5}{d\tilde{t}_s} &= F_5(v_0, v_1, v_2, v_3, y_5), \\
\frac{dv_1}{d\tilde{t}_s} &= F_1(y_6), \\
\frac{dy_6}{d\tilde{t}_s} &= F_6(v_0, v_1, y_6), \\
\frac{dv_2}{d\tilde{t}_s} &= \varepsilon F_2(y_7), \\
\frac{dy_7}{d\tilde{t}_s} &= \varepsilon F_7(v_0, v_2, y_7),
\end{aligned}
\tag{7}
$$

which describes the dynamics on $S^0$. System (7) is a two-time-scale problem for $\varepsilon$ sufficiently small and it gives the *slow layer problem* in the $\varepsilon \to 0$ limit,

$$0 = F_3(y_8),$$
$$0 = F_8(v_0, v_2, v_3, y_8),$$
$$\frac{dv_0}{d\tilde{t}_s} = F_0(y_5),$$
$$\frac{dy_5}{d\tilde{t}_s} = F_5(v_0, v_1, v_2, v_3, y_5),$$
$$\frac{dv_1}{d\tilde{t}_s} = F_1(y_6),$$
$$\frac{dy_6}{d\tilde{t}_s} = F_6(v_0, v_1, y_6),$$

(8)

where $v_2$ appears as a parameter. A periodic orbit of (7) for $\varepsilon = 0.01$ and the bifurcation diagram of (8) as a function of $v_2$ is projected on the $(v_0, v_2)$-plane in **S1** Fig(b).

In the slow layer problem (8), the slow variables $(v_0, y_5, v_1, y_6)$ evolve along fibers defined by $(v_3, y_8, v_0, y_5, v_1, y_6, v_2, y_7) = (v_3, y_8, v_0, y_5, v_1, y_6 v_2^0, y_7^0)$, where $(v_2^0, y_7^0)$ are constant and $(v_3, y_8, v_0, y_5, v_1, y_6 v_2^0, y_7^0)$ restricted to $S^0$. The equilibria of (8) defines the *superslow manifold* $L^0$

$$L^0 = \{(v_3, y_8, v_0, y_5, v_1, y_6 v_2, y_7) \in S^0 \mid \begin{array}{l} F_3(y_8) = 0 \ \cap F_8(v_0, v_2, v_3, y_8) = 0 \ \cap \\ F_0(y_5) = 0 \ \cap \ F_5(v_0, v_1, v_2, v_3, y_5) = 0 \ \cap \\ F_1(y_6) = 0 \ \cap \ F_6(v_0, v_1, y_6) = 0 \end{array} \},$$

which is a subset of $S^0$. The superslow manifold $L^0$ is reduced to

$$L^0 = \{(v_3, y_8, v_0, y_5, v_1, y_6 v_2, y_7) \in S^0 \mid$$
$$A\,S\big(A\tau_a p + C_2\tau_a A\,S(C_1\tau_a v_0) - C_4\tau_b v_2 - C_7\tau_g v_3\big) - v_0 = 0\},$$

where $v_3 = G\,S(C_5\tau_a v_0 - C_6\tau_b v_2)$ is on $S^0$. The curve $L^0$ perturbs to locally slow invariant manifolds for $\varepsilon > 0$ along the hyperbolic branches of $L^0$, while the dynamics of near the non-hyperbolic fold points should be investigated using GSPT. Finally, the *superslow reduced problem* obtained by setting $\varepsilon \to 0$ in (5) reads

$$\frac{dv_2}{d\tilde{t}_s} = F_2(y_7),$$
$$\frac{dy_7}{d\tilde{t}_s} = F_7(v_0, v_2, y_7). \tag{9}$$

This algebraic-differential system determines the superslow dynamics restricted to $L^0$, and eventually to $S^0$.

**Clinical data**

Clinical data used for the purpose of this study consisted in SEEG signals collected in a patient with drug-resistant focal epilepsy that required invasive EEG exploration. Recordings were performed using intracranial multichannel electrodes (DIXI Medical, 5–18 contacts; length, 2 mm, diameter, 0.8 mm; 1.5 mm apart). Electrodes were implanted according to the stereotactic method of Talairach [22]. SEEG signals were recorded on a Deltamed™ system on a maximum number of channels equal to 128, and were sampled at 256 Hz and recorded to hard disk (16 bits/sample) using no digital filter. The only filter present in the acquisition procedure was a hardware analog high-pass filter (cut-off frequency of 0.16 Hz) used to remove very slow variations that sometimes contaminate the baseline. In the patient for which data is displayed in the remainder of the manuscript, a surgical operation was performed 6 month after pre-surgical exploration (cortectomy of the frontal dorsolateral region). Histological data revealed the presence of a focal cortical dysplasia (Taylor). After surgery, the patient was seizure free (Engel IA). As a reminder, SEEG is always carried out as part of normal clinical care of patients who give informed consent in the usual way. Patients were informed that their data may be used for research purposes.

**Computational methods**

The bifurcation analysis was in done with AUTO-07p [23]. Model equations were implemented in XPPaut [24]. Stochastic differential equations were iterated using Euler-Maruyama method with a step size $dt = 10^{-4}$ second.

## Results

### Pre-ictal spiking during interictal to ictal transition

In partial (i.e. focal) epilepsies, the onset of seizures is characterized by the appearance of a rapid discharge, typically in the gamma frequency band ([25, 120] Hz) [25]. This fast onset activity has long been recognized as a hallmark of the epileptogenic zone, and a number of methods have been proposed to make use of this pattern to identify the epileptogenic zone [26,27]. Interestingly, fast onset activity is often preceded by a specific electrophysiological pattern consisting of sustained large amplitude bursts with superimposed faster spikes, which can be observed in various etiologies [28]. A typical example of this pre-ictal pattern, as recorded in a patient during pre-surgical investigation with depth electrodes, is shown in **Fig 1**. As depicted, this dynamical regime starts with sporadic bursts, which become periodic to change into a sustained discharge of pre-ictal bursts. In this case, the number of spikes of the bursts gradually decreases during the pre-ictal burst phase, which continues approximately for 14 seconds. The pre-ictal burst phase is followed by the fast activity that actually marks the onset of the seizure.

**Fig 1. SEEG signals recorded in a patient with epilepsy during the interictal to ictal transition and simulated signals**. (a) Epileptic seizure recorded in a patient showing the typical pre-ictal spiking pattern with three phases: sporadic spikes, pre-ictal bursts, and fast onset. (b) Zoom into each phase of the actual SEEG signal. The pre-ictal burst type-1 is followed by the pre-ictal burst type-2. (c) Simulated signals corresponding to each phase.

System (1) represents a physiologically relevant system, and has been extensively explored to establish relationships between model parameters and electrophysiological patterns observed in SEEG recordings [16,29]. For instance, increasing the ratio of the synaptic gain of the excitatory pyramidal cell population and inhibitory SOM+ interneuron population introduces a region in the parameter space where the system can undergo different stages of epileptogenic activity as a function of the synaptic gain of inhibitory SOM+ interneuron population, parameter $B$, and the synaptic gain of inhibitory PV+ interneuron population, parameter $G$.

The thorough exploration of system (1) led to the identification of three key parameters: $B$, $G$, and the strength of the excitatory synaptic coefficient on the PV+ interneuron subpopulation $C_5$. Indeed, the tuning of these three parameters enables replicating of the different pre-ictal stages shown in **Fig 1**a. These results are illustrated by the bifurcation diagram in **Fig 2**a. As depicted, the decrease of parameter $B$ yields a transition from background activity to fast onset activity, though pre-ictal spiking. **Fig 2**b shows where these activity regions are localized in the $(B, G)$-parameter space.

**Fig 2. Bifurcation diagrams of the system** (1) **showing the different pre-ictal stages numbered 1-3 in Fig 1.** (a) Amplitude of the PSP of the pyramidal cell subpopulation is plotted as a function of the synaptic gain of the SOM+ interneuron subpopulation $B$. Bold and dashed lines correspond to stable and unstable solutions, respectively. Region 1 (blue) corresponds to sporadic bursts, region 2 (orange) to sustained bursts, and region 3 (purple) to low voltage fast onset activity. The system yields large amplitude ∼ 30 Hz oscillations in the unnumbered green shaded region. The unnumbered gray shaded area corresponds to high amplitude stable equilibrium points. The arrow shows the route from background to low voltage fast onset activity in the parameter space. (b) Co-dimension 2 diagram of the Hopf (H) and LP bifurcation points marked on panel (a) in the parameter space of $B$ and $G$ (synaptic gain of the PV+ interneuron subpopulation). The $LP_1$ and $LP_2$ points merge on a cups (CP) bifurcation, and the $H_1$ and $H_2$ merge on a zero-Hopf (ZH) bifurcation.

Let us walk through the bifurcation diagram in **Fig 2**a, starting the equilibrium point at $B = 0$, and increase $B$. The system first undergoes a Hopf bifurcation ($H_1$) at $B \approx 1.44$, giving stable limit cycles at $\approx 30$ Hz (gamma activity). The amplitude of these solutions increases with $B$. These oscillations terminate on another Hopf bifurcation ($H_2$) at $B \approx 4.55$. Then, we identify a third Hopf bifurcation ($H_3$) for $B \approx 7.74$, where a branch of stable oscillations around $\approx 6$ Hz appears. As B increases, this branch connects to stable bursting orbits by passing through several limit folds along the vertical zigzags around $B \approx 8.86$. At $B \approx 10$ the system reaches to the maximum number of spikes per burst orbit (11 spikes for this parameter set). Increasing $B$ decreases the number of spikes *via* the horizontal zigzags in $y_0$ between $B \in (9.5, 22.5)$. The bursts terminate at $B \approx 22.5$. The branch holding the unstable equilibrium points forms a Z-shaped curve with two folds (limit points (LP)) at $B \approx 21.3$ (LP$_3$) and $B \approx 35.6$ (LP$_4$), with unstable focus on the upper branch, saddles in the middle and stable nodes on the lower branch after a subcritical Hopf bifurcation ($H_4$) which gives unstable limit cycles making a heteroclinic connection with the middle branch. For $B > 35.6$, the system only has stable equilibrium points as solutions.

Continuation of the LP and H points marked on **Fig 2**a in the $(B, G)$-space is shown in **Fig 2**(b). It can be seen that the locations of LP$_3$, LP$_4$, H$_3$, H$_4$ points do not depend on $G$, whereas the locations of LP$_1$, LP$_2$, H$_1$ and H$_2$, which are related to the oscillations at $\approx 30$ Hz, do. The fast onset region does not exist for small values of B if $G < 5$. Furthermore, $G$ controls the amplitude of the spikes of bursting solutions, which increases with $G$.

Assume that system (1) is initially in the background activity mode, which corresponds to the white region in **Fig 2**a for $B > 35.6$, where unique stable equilibrium points on the bifurcation curve is observed. In the blue region between the two folds LP$_3$ and LP$_4$, the bifurcation curve takes a Z-form with stable nodes on the lower branch, unstable nodes in the middle and saddle-nodes on the upper branch. For $B$ values in this blue region, system (1) under a stochastic input $p(t)$ undergoes sporadic bursts, with an increasing probability as $B$ approaches to the left fold. As $B$ decreases, the system enters into the bursting region (orange region). Note that further increasing B would increase the number of spikes. However, in the recordings we see that the number of spikes decreases in the course of the pre-ictal bursting regime as the system approached the low voltage fast onset (LVFO), transition from type-1 to type-2 bursting. This change is very subtle to be reproduced in the model because, as detailed in the Sec. *Burst analysis,* the number of spikes depends on the presynaptic potential on PV+ interneurons: the lower it is, the more spikes within the burst are obtained. Thus, the number of spikes increases when the inhibitory input decreases, or when the excitation onto PV+ interneurons increases. At this stage, transition from the type-1 bursting to type-2 bursting is obtained by keeping $B$ constant, but decreasing the excitatory post-synaptic potential (EPSP) on PV+ by decreasing progressively $C_5$ to 300 to reduce the number of spikes; and increase $G$ to increase spikes amplitude. Under these variations, the bifurcation diagram in **Fig 2**a remains qualitatively the same, the most important quantitative change being the location of the Hopf bifurcation point $H_1$ related to the LVFO. As shown in **Fig 2**b, increasing $G$ moves the $H_1$ towards right along the $B$-axis, and initiates the LVFO for low values of $B$.

**Bursting analysis**

We investigate the bursting dynamics of system (1) using system (3), which is a kind of nondimensionalized version of (1) but expressed in an explicit slow-fast formulation. **Fig 3**a shows a bursting solution of (3) in the $(v_0, v_2, v_3)$-space, the critical manifold $S^0$ and the superslow manifold $L^0$ (see Sec. *Slow-fast formulation of the model* for definition). The critical manifold $S^0$ is normally hyperbolic (not folded) and stable, and stretches between almost horizontal surfaces (lower and upper) with an almost vertical plane. The superslow manifold $L^0$ has branches both on the lower horizontal surface and vertical surface of $S^0$. While the part of $L^0$ on the vertical surface of $S^0$ is stable, the part on the lower horizontal surface of $S^0$ is divided into stable and unstable sections at two fold points $LP_1$ and $LP_2$. The curve $L^0$ is stable along the branch that is almost parallel to the $v_2$-axis, unstable along the branch between $LP_1$ and $LP_2$, and then becomes stable again. The stable and unstable branches of $L^0$ are normally hyperbolic, whereas the fold points $LP_1$ and $LP_2$ are not.

The critical manifold $S^0$ and superslow manifold $L^0$ perturb for small enough values of time-scale parameters, hence the three-time-scale dynamics of (3) approximate to $S^0$ and $L^0$. During the superslow time-scale under (9), the bursting orbit follows the stable branch of $L^0$ almost parallel to the $v_2$-axis. Near the fold point $LP_1$, the trajectory bends in the $v_3$-direction along the vertical plane of $S^0$ and enters into the spiking regime, which runs in fast time-scale. The spiking terminates close to $LP_2$ and the trajectory jumps back to the stable branch of $L^0$ almost parallel to the $v_2$-axis in slow time-scale under (8). **Fig 3**b shows the time series in $\tilde{t}$ of the orbit in **Fig 3**a.

**Fig 3. Bursting orbit of system** (3)**.** (a) Solution of (3) (blue orbit) and $L^0$ (red curve) on the critical surface $S^0$(green surface) projected on the $(v_0, v_2, v_3)$-space. Single-headed, double-headed and triple-headed arrows indicate the flow direction during superslow, slow and fast time-scales, respectively. LP denotes limit point bifurcation. The $L^0$ curve changes stability at two limit points, $LP_1$ and $LP_2$ (red dots). The middle branch of the $L^0$ curve between these limit points is unstable (dashed). (b) Time course of the variables $(v_3, v_0, v_2)$ of the orbit plotted in panel (a). (c) Solution of (3) projected on the bifurcation diagram (black curve) of (4) for ε=0 where $v_2$ is threaded as a parameter. Arrows show the direction of the flow with respective time-scales. Bold and dashed lines correspond to stable and unstable solutions, respectively. H donates a Hopf bifurcation, LP a limit point bifurcation. The equilibrium

points along the black Z-shaped curve are unstable on the middle branch of the curve, between $LP_1$ at $v_2^{LP_1} = 4.778$ and $LP_2$ at $v_2^{LP_2} = 20.66$ (black dots), and on the upper branch between H$_1$ at $v_2^{H_1} = 0.27$ and H$_2$ at $v_2^{H_2} = 14.27$ (green dots). The amplitude of the stable limit cycles is bounded by the green continuous curves connecting the $H_1$ and $H_2$ in the ε=0 limit.

For a better understanding of the bursting dynamics, we consider system (4) at $\varepsilon = 0$ for which the variables of the slowest subsystem $(v_2, y_7)$ act as parameters of the $(v_3, y_8, v_0, y_5, v_1, y_6)$-subsystem. Since only $v_2$ appears in the $(v_3, y_8, v_0, y_5, v_1, y_6)$-subsystem, its dynamics depend on $v_2$. In **Fig 3**c, the bursting orbit in **Fig 3**a is superimposed on the bifurcation diagram of the $(v_3, y_8, v_0, y_5, v_1, y_6)$-subsystem in (4) at $\varepsilon = 0$ as a function of $v_2$. Although the fastest variables of (4) are $(v_3, y_8)$, we chose $v_0$ vs $v_2$ for a clearer visualization (the same trajectory and the bifurcation diagram are given on the $(v_3, v_2)$-plane **Fig 4**). We see that the corresponding system poses a Z-shaped bifurcation diagram as a function of $v_2$ with two folds, $v_2^{LP_1}$ and $v_2^{LP_2}$. The equilibrium points are stable on the lower branch of the Z-shaped curve for $v_2 > v_2^{LP_1}$, unstable along the middle branch between $v_2^{LP_1}$ and $v_2^{LP_2}$. The upper branch has two supercritical Hopf bifurcations, at $v_2^{H_1}$ and $v_2^{H_2}$, with stable limit cycles in between. Along the upper branch, equilibrium points are stable for $v_2 < v_2^{H_1}$ and $v_2^{H_2} < v_2 < v_2^{LP_2}$. The bursting behavior resulting from this bifurcation structure in the $(v_3, y_8, v_0, y_5, v_1, y_6)$-subsystem is classified as 'fold/Hopf bursting' by Izhikevich [30] due to the presence of a 'fold/Hopf' hysteresis in the bifurcation diagram.

System (4) may undergo through these bifurcations in a repetitive manner for $\varepsilon \neq 0$, which results eventually in the bursting solutions for small enough values of $\varepsilon$. As the arrows on **Fig 3**c and the traces on **Fig 3**b demonstrate, the trajectory follows the lower stable branch during the quiescence phase of the bursting, which terminates near $v_2 \approx v_2^{LP_1}$. Then, it jumps to the region of the stable limit cycles on the upper branch, which initiates the active phase of the bursting. The spiking frequency during the active phase is faster at the beginning than the end due to the fact that the Hopf bifurcation at $v_2^{H_1}$ gives limit cycles with $\approx 30$ Hz frequency whereas the Hopf bifurcation at $v_2^{H_2}$ gives limit cycles with $\approx 10$ Hz. The spiking terminates at $v_2 \approx v_2^{H_2}$, but the active phase continues until the trajectory jumps back to the stable lower branch at $v_2 \approx v_2^{LP_2}$. We underline that as $\varepsilon \to 0$, the bursting orbit attaches more and more the bifurcation diagram obtained for $\varepsilon = 0$ (see **S1 Fig** for an example).

The main difference between the type-1 and type-2 bursting is the number of spikes during the active phase of bursting. In the model, the variations in the number of spikes can be met by changing the excitation level on the PV+ interneurons: as aforementioned, the number of spikes increases with the amount of excitation received by PV+ interneurons. This can be achieved by either decreasing inhibition or by increasing excitation. For instance, decreasing $B$ in region-2 in **Fig 2** increases the number of spikes. In (4) at $\varepsilon = 0$, the excitation on PV+ depends on two synaptic coupling coefficients, $C_5$ and $C_6$. The effect of $C_6$ will be similar to the one of $B$, since they both scale the PSP of SOM+ interneurons given by the variable $v_2$ in (4). Below, the role of the excitatory synapses in the $(v_3, y_8, v_0, y_5, v_1, y_6)$-subsystem by changing $C_5$ is investigated.

As displayed by **Fig 3**c, **Fig 4**a and **Fig 4**b, the spikes are bounded by $LP_1$ and $H_2$ in the bifurcation diagram of the $(v_3, y_8, v_0, y_5, v_1, y_6)$-subsystem as a function of $v_2$. The distance between $LP_1$ and $H_2$ in $v_2$ affects the number of spikes; the further they are, the more spikes the burst has. In **Fig 4**c, LP and Hopf bifurcations are continued in the parameter space of $(v_2, C_5)$. While the $LP_1$ and $LP_2$ lie along almost vertical lines, the Hopf bifurcation points form a V-shaped curve along which the left arm locates the $H_1$ points and the right arm the $H_2$ points. The distance between $H_2$ and $LP_1$ increases with $C_5$, hence, the spike number. At $C_5 = 139$, $H_2$ and $LP_1$ are aligned. Further decrease in $C_5$ places $H_2$ on the left of $LP_1$ and leaves no chances for a bursting solution. The system yields only relaxation type of oscillations for $C_5 < 139$.

**Fig 4. Spike number as a function of $C_5$.** (a) Solution of (4) with 3 spikes for $C_5 = 500$ projected on the bifurcation diagram of the fast system (5) as a function of $v_2$. Arrows indicate the direction of the flow. (b) Solution of (4) with 1 spike for $C_5 = 300$ projected on the bifurcation diagram of the fast system (5) as a function of $v_2$. Arrows indicate the direction of the flow. (c) Co-dimension 2 diagrams of the Hopf (H) points (green) and the LP points (black) in the $(v_2, C_5)$ parameter space marked on the left and middle panels. As $C_5$ decreases, $H_2$ moves leftwards and eventually the spike number decreases. For $C_5 = 139$, $H_2$ and $LP_1$ are aligned at $v_2 = 4.778$. A further decrease in $C_5$ places $H_2$ on the left of $LP_1$ and leaves no chances for a bursting solution.

Overall, the aforementioned analysis shows that pre-ictal bursting runs in three-time-scales. The system sustains the bursting regime for a certain range of parameter $B$ denoting SOM+ synaptic gain. The complex pre-ictal bursting pattern can be accurately adjusted by tuning parameters $G$, which controls the PV+ synaptic gain, and the connectivity coefficient $C_5$, which controls PV+ excitability. In particular, the number of spikes and their amplitude can be adjusted by tuning $C_5$ and $G$, respectively.

**Perturbation analysis**

Electrical (through direct stimulation) and optical (through optogenetics, using light pulses in genetically modified animals) perturbations can alter action potential firing through modification of the mean membrane potential of the considered neural subpopulation. We assumed an additive model for the stimulation effect onto the mean membrane PSP [31].

Thus, the external input $I_{ext}(t)$ is included in the 'pulse-to-wave' functions of the neural mass model under the following form:

$$S\left(k_i\, I_{ext}(t) + \sum I_{i,PSP}\right)$$

where $k_i$ is the coupling coefficient between the stimulation and the considered subpopulation, governing the impact of $I_{ext}(t)$ on the subpopulation, $\sum I_{i,PSP}$ is the total afferent received by the subpopulation $i = \{P, SOM, PV\}$.

A pulse input (biphasic or monophasic) changes the PSP of the perturbed subpopulation by shifting it above its base level $S(0)$. We assume that a neural mass block, given by $\ddot{y} = W/\tau_w\, S\left(I_{ext}(t)\right) - 2/\tau_w \dot{y} - 1/\tau_w^2\, y$, receives biphasic stimulation. The PSP of the neural mass block increases during the anodal pulse, but decreases (discharges) between the inter-pulse intervals of the biphasic input. Depending on the pulse width, pulse amplitude, and mostly on the synaptic time constant of the neural mass block, this shift may be sustained or not. For instance, the discharge takes longer in a neural mass block with slow synaptic kinetics than the one with fast synaptic kinetics. If the pulse frequency is sufficiently high to stimulate the neural mass before it completely resumes to its base level, then the PSP of the neural mass can oscillate above the base level. As visualized in **S2** Fig, the same perturbation shifts the PSP of a neural mass with slow synaptic kinetics, while the subsystem with fast synaptic kinetics decays to its base level during the inter-pulse intervals of the stimulation. Increasing the stimulation frequency can keep the PSP of the subsystem with fast kinetics above the base level, and therefore the firing rate and PSP of the fast subsystem increase with the stimulation frequency.

The bursting solution is driven by the slow oscillations in system (3) (see Sec. *Bursting Analysis*). The slow dynamics of (3) (subsystems representing the pyramidal cell and SOM+ interneuron subpopulations) can be approximated by (7), which preserves the burst-driver slow oscillations behavior for the same parameter values yielding bursting oscillations in (3) (see Sec. *Slow-fast formulation* and **S1** Fig). Thus, it is sufficient to investigate the response of (7) under perturbation to understand the impact of the perturbation on the burst-driver slow dynamics. The most common signal delivered to brain tissue in the field of DBS is bi-phasic pulses with balanced anodic/cathodic phases of brief durations (approximately 100 µs). Below, the impact of anodic and cathodic constant external inputs is considered without taking into account their duration, to simply understand how they alternate the phase space of system (7). For this purpose, a constant input ($I_{ext} = 1$) scaled with the impact coefficients $k_P$ and $k_{SOM}$ is applied to (7).

In **Fig 5**, subpopulations representing pyramidal cells and inhibitory SOM+ interneurons are perturbed. The left panels of **Fig 5** show the $y_5$-nullsurface $\Theta$, $v_2$-nullsurface $\Sigma$ and the superslow manifold $L^0$ projected on the $(y_5, v_2, v_0)$-space. The solution of (7) is visible on the left panels, and the solution of (3) for the same parameters is given on the right panels of **Fig 5**. **Fig 5**a shows the case where only the SOM+ interneuron subpopulation described by the $(v_2, y_7)$-subsystem in (7) is subject to the constant external input ($k_P = 0$). In the absence of any perturbation ($k_{SOM} = 0$), $\Theta$ and $\Sigma$ intersect for $v_2 > 20$. System (7) has a limit cycle which flows on $\Theta$ and (3) a burst orbit (black solutions **Fig 5**a and **Fig 5**a1, respectively). The quiescence phase of the burst corresponds to the slow passage following $L^0$ where $v_0 \approx 0$, and the active phase correspond to the trajectory on the upper sheet of $\Theta$.

**Fig 5. Geometrical analyses of a constant input.** Constant input is applied to SOM+ interneurons (a), to pyramidal cells (b) and to both SOM+ interneurons and pyramidal cells (c). Left panels show the projection of the nullsurfaces, critical slow manifold and the orbit of the reduced model (7). Right panels show the LFP signal of the full system (3) subject to the constant inputs analyzed on the left. All parameters are as given in Table 1, except $B = 15$.

(a) The $y_5$-nullsurface $\Theta$ (blue surface for $k_P = 0$), and $y_7$-nullsurface $\Sigma$ (red surface for $k_{SOM} = -1$, black surface for $k_{SOM} = 0$, green surface for $k_{SOM} = 1$) are projected on the $(v_2, y_5, v_0)$-space. The blue curve $L^0$ (stable on the bold, unstable on the dashed) is on the intersection between $\Theta$ and the $\{y_5 = 0\}$-hyperplane. The black and red orbits are the solutions of the system for $k_{SOM} = 0$ and $k_{SOM} = -1$, respectively. For $k_{SOM} = 1$, the solution approaches to the green stable equilibrium point on the intersection between $\Sigma(k_{SOM} = 1)$ and $L^0$. Panel (a1) shows the time series for $k_{SOM} = \{0, 1\}$, and panel (a2) for $k_{SOM} = -1$. (b) The $y_5$-nullsurface $\Theta$ (red surface for $k_P = 1$, green surface for $k_P = -1$), and $y_7$-nullsurface $\Sigma$ (black surface for $k_{SOM} = 0$) are projected on the $(v_2, y_5, v_0)$-space. The red curve $L^0$ (stable on the bold, unstable on the dashed) is on the intersection between $\Theta(k_P = 1)$ and the $\{y_5 = 0\}$-hyperplane. The green curve $L^0$ (stable on the bold, unstable on the dashed) is on the intersection between $\Theta(k_P = -1)$ and the $y_5 = 0$ hyperplane. The green and red orbits are the solutions of the system for $k_P = 1$ and $k_P = -1$, respectively. Panel (b1) shows time series for $k_P = 1$, and panel (b2) for $k_P = -1$. (c) The $y_5$-nullsurface $\Theta$ (red surface for $k_P = 1$) and $y_7$-nullsurface $\Sigma$ (green surface for $k_{SOM} = 1$, blue surface for $k_{SOM} = 2$) are projected on the $(v_2, y_5, v_0)$-space. The red curve $L^0$ (stable on the bold, unstable on the dashed) is on the intersection between $\Theta(k_P = 1)$ and the $y_5 = 0$ hyperplane. The green curve $L^0$ (stable on the bold, unstable on the dashed) is on the intersection between $\Theta(k_P = 1)$ and the $\{y_5 = 0\}$-hyperplane. The green orbit is the solution of the system for $(k_P, k_{SOM}) = (1, 1)$. For $(k_P, k_{SOM}) = (1, 2)$ the solution approaches to the cyan stable equilibrium point on the intersection between $\Sigma(k_{SOM} = 2)$ and $L^0$. Panel (c1) shows time series for $(k_P, k_{SOM}) = (1, 1)$, and panel (c2) for $(k_P, k_{SOM}) = (1, 2)$.

A key point in terms of controlling bursting activity through direct stimulation is that an input leading to a bifurcation from the stable limit cycle to an equilibrium point can prevent the system from bursting by keeping the system in the silent phase. This can be achieved by an input that ensures an intersection between $\Sigma$ hyperplane and the lower branch of $L^0$. Indeed, for $k_{SOM} = 1$, (7) possesses a stable equilibrium point near the left fold of $L^0$ which traps the trajectory (green dot in **Fig 5**a). For the same input, the bursting in (3) is aborted (green solution in **Fig 5**a1). On the other hand, a negative constant input ($k_{SOM} = -1$), moves $\Sigma$ away from the left fold of $L^0$. Being $\Sigma$ closer to the upper branch of $L^0$ prolongs the active phase of the burst and increases the number of spikes, as seen in **Fig 5**a2. These observations indicate that increasing the excitation on SOM+ interneurons can abort bursting.

In **Fig 5**b, only the subsystem representing the pyramidal cells receives the perturbation ($k_{SOM} = 0$). The input on the pyramidal cell subpopulation acts on $\Theta$. While positive constant input ($k_P = 1$) increases the distance between the lower fold of $L^0$ and $\Sigma$, negative constant input ($k_P = -1$) decreases this distance. Both systems (7) and (3) preserve the oscillatory behavior for these values of $k_P$, yet, the oscillation frequency decreases for $k_P = -1$ due to the decreased distance between the lower fold of $L^0$ and $\Sigma$). Thus, hyperpolarization of pyramidal cells by increasing inhibition on them can abort bursting.

As aforementioned above, pulsed stimulation increases the firing rate of a neuronal population. However, a stimulation applied to one specific region might not affect all neural populations in the same manner. This can be due to the relative position of electrodes with respect to neurons, cell specific firing thresholds, or synchronization level within neural subpopulations. However, such features can bring certain advantages in aborting bursting. **Fig 5**c shows the response of the system when both subpopulations of pyramidal cells and SOM+ interneurons are perturbed, the oscillatory behavior in systems (7) and (3) continues under the same positive constant input ($k_P = k_{SOM} = 1$). With such input, the number of spikes during the active phase is decreased (**Fig 5**c1). If the subpopulation of SOM+ interneurons is perturbed more strongly than the subpopulation of pyramidal cells ($k_P = 1, k_{SOM} = 2$), the system can bifurcate to the resting state (**Fig 5**c2).

Although the reduced system (7) does not include the fast dynamics of the PV+ interneurons, the effect of the perturbation on the subpopulation of PV+ interneurons can be understood geometrically. First, let us notice that increasing the inhibition on SOM+ interneurons encourages spiking (**Fig 5**a, **Fig 5**a2), while increasing the excitation on SOM+ aborts bursting (**Fig 5**a, **Fig 5**a1). Perturbing (stimulating) the subpopulation of PV+ interneurons increases the PSP from PV+ interneurons to pyramidal cells, reduces the PSP from pyramidal onto SOM+, and in turn favors bursting. Another way to illustrate the impact of perturbing the PV+ on bursting is to examine the diagram in **Fig 4**. Anodic pulses (positive perturbation or depolarization of the membrane potential) can shorten the quiescent phase in the 'fold/Hopf' hysteresis loop by kicking the trajectory to the region of stable limit cycles between the two Hopf bifurcations $H_1$ and $H_2$. Hence, such pulses applied periodically can increase the bursting frequency by shortening the quiescent phase. On the other hand, cathodic pulses (negative perturbation or hyperpolarization of the membrane potential) can lengthen the quiescent phase by hooking the trajectory near the down state of the hysteresis loop.

Overall, this geometric perturbation analysis helps to clarify the role of hyperpolarizing and depolarizing inputs on ongoing bursting activity. In particular, depolarization of the subpopulation of SOM+ interneurons or hyperpolarization of the subpopulation of pyramidal cells can abort bursting by keeping the sum of PSPs at low levels. Depolarization of the subpopulation of PV+ interneurons contributes to bursting.

**Stimulation applied during the pre-ictal burst regime**

The analysis in Sec. *Perturbation Analyses* has shown that a positive constant input applied to the subpopulation of SOM+ interneurons can bifurcate the limit cycle (oscillating epilepsy-like activity) to an equilibrium point (background activity), while a positive constant input on the subpopulations of pyramidal cells and PV+ interneurons preserve bursting and high frequency oscillations (**Fig 5**). Hence, an appropriate strategy for pre-ictal bursting abortion consists in the excitation of the SOM+ interneuron subpopulation.

In this section, the results obtained from the mathematical analysis are translated into an *in silico* set-up mimicking experimental conditions. Typically, charge-balanced bi-phasic pulses (pulse width = 0.5 ms and total duration 1 ms) with an arbitrary unit (arb. unit) amplitude are applied during the pre-ictal bursting/spiking regime in the presence of a stochastic input. In order to test our predictions on the role of different neural populations, only SOM+ interneurons are perturbed in **Fig 6**a ($k_{SOM} = 1, k_P = 0, k_{PV} = 0$), whereas in **Fig 6**b all neural subpopulations are perturbed homogenously (coupling coefficients $k_{SOM} = k_P = k_{PV} = 1$).

Results indicates that pre-ictal bursts frequency decreases when the stimulation is switched on at, typically at the instant $t = 5s$ in both cases. The bursting regime can be aborted if the stimulation frequency and amplitude are sufficiently high. The minimum values of the stimulation frequency and amplitude to abort bursting depend on which neuronal subpopulation receives the stimulation. When only the SOM+ interneuron subpopulation is stimulated, the minimum stimulation frequency and amplitude required to abort bursting are lower than the case where all neural subpopulations are stimulated homogenously. As exemplified in **Fig 6**c, bursting is suppressed at f = 15 Hz for an amplitude of 10 arb. unit when only SOM+ interneurons are stimulated. While the same stimulation can considerably decreases the frequency of bursting events (**Fig 6**d) when all subpopulations are impacted, the stimulation frequency should be increased to 25 Hz for a complete bursting suppression (**Fig 6**e).

The difference between type-1 and type-2 bursting is the number of spikes during the active phase that is related to the excitatory input onto PV+ interneurons. In particular, the EPSP is larger in the former case. Despite this difference, the bursting mechanisms in both types are the same; i.e. slow oscillations in the SOM+ interneurons drive sequentially and periodically the same type of bifurcations in the subsystem of pyramidal cells and PV+ interneurons. Hence, the strategy for aborting bursting relying on aborting oscillations in the SOM+ subsystem does not depend on the bursting type. The estimations on the stimulation parameters (in terms of frequency and amplitude) given in **Fig 6**c, which are for type-1 bursting, are capable of aborting type-2 bursting and sporadic bursting, as well, because both of the regimes are less excited than type-1 bursting.

**Fig 6. System** (1) **under stimulation.** Biphasic stimulation with a 0.5 ms pulse width (total pulse duration is 1 ms) is applied to the system in type-1 bursting. Panels (a) and (b) show the energy map of the simulated LFP signal is lower in the blue region than the yellow region (see the color bar on the right). (a) Only the SOM+ interneurons receive the biphasic perturbation. (b) The pyramidal cell, SOM+ interneurons and PV+ interneurons receive the same biphasic perturbation. Panels (c), (d) and (e) show the time course of the marked stimulation on panels (a) and (b). (c) 15 Hz biphasic stimulation with 10 arb. unit amplitude is applied to the SOM+ interneurons ($k_{SOM} = 1, k_P = k_{PV} = 0$). (d) 15 Hz biphasic stimulation with 10 arb. unit amplitude is homogenously applied to all subpopulations ($k_{SOM} = k_P = k_{PV} = 1$). (e) 25 Hz biphasic stimulation with 10 arb. unit amplitude is homogenously applied to all subpopulations ($k_{SOM} = k_P = k_{PV} = 1$).

## Discussion

Epilepsy is a dynamic and complex disease running on different time-scales [32,33]. Epileptic activity is characterized by long interictal periods (outside seizures), during which the brain behaves mostly as a normal brain, then marked by brief ictal episodes (seizures). The seizure onset, i.e. the transition from interictal to ictal activity, has a wide repertoire in human focal epilepsies [28,34]. In this study, we focused on a specific electrophysiological pattern generally referred to as "pre-ictal spikes" or "pre-ictal discharges", which has been particularly described in mesial temporal lobe seizures [11–13] but that may also be observed as a seizure onset pattern in neocortical seizures from various origins [28,35]. This complex pattern is signed by large amplitude fast spikes followed by a slow discharge, thus holding the properties of a bursting and is called "pre-ictal bursting" in this paper.

We successfully reproduced the complex pre-ictal bursting pattern in a NMM featuring three subsets of neurons (subpopulations of pyramidal neurons, SOM+ and PV+ interneurons) in [16]. The slow-fast formulation of the model unveiled its three-time-scale structure and the following analysis detailed the mechanisms responsible for the pre-ictal bursting. In particular, the bursting process in the model arose from a high level of excitation among pyramidal neurons as well as onto the PV+ interneuron subpopulation. In the bursting regime, the slow oscillations mediated by the SOM+ interneurons are the drivers of bursting solutions, and the number of spikes during an active phase of a burst depends on the level of excitation on the PV+ interneurons. Ultimately, we showed that a perturbation that was able to stop the slow oscillations in the SOM+ interneuron subpopulation would be sufficient to stop pre-ictal bursting activity.

These model predictions corroborate some experimental findings. Indeed, *in vitro* data from human specimen suggested that a glutamate-dependent cellular and/or synaptic plasticity process underlies the occurrence of pre-ictal discharges during the transition to seizure. Pre-ictal discharges would initiate changes in glutamatergic and GABAergic signaling in groups of neurons larger than those involved in interictal discharges. Repeated discharges would result from a dynamic process that ultimately leads to ictal events [36]. Along the same line, as extensively reviewed in [37], both excitatory and inhibitory networks are involved in epileptogenesis and seizure generation. In particular, GABAergic-mediated mechanisms contribute to synchronizing neuronal networks in epileptic brain structures. Notably, interneuronal activity is enhanced and synchronized during sustained epileptic spikes [38,39].

This viewpoint is particularly interesting if the role of the GABAergic system in the suppression of epileptiform pre-ictal activity is considered when direct brain stimulation applied during the interictal period. For instance, optogenetic stimulation of the CA3 region of hippocampus leads to considerable reduction of seizures in the CA3 region by entrainment of GABAergic interneurons targeting $GABA_A$ receptors [40,41]. Low-frequency stimulation of fiber tracts during the inter-ictal period has also been shown to reduce seizures through activation of the $GABA_B$ signaling in animal models of temporal lobe epilepsy activity [42–44], as well as with the application of an electrical field [45]. The success of low-frequency stimulation of fiber tracts in focal cortical seizures has also been linked to GABAergic effects [46,47].

Our results are in line with the above reported data, and indicate that an abortive stimulation of the epileptic activity during the pre-ictal bursting regime should primarily target the GABAergic system (mostly on interneurons with slow synaptic kinetics). Stimulating the GABAergic system yielded more pronounced effect as compared with the stimulation pyramidal neurons. The stimulation frequency required to change the PSPs of neural subpopulations was directly linked with their kinetics: the slower they are, the lower stimulation frequency needs to be. At this point, SOM+ interneurons were impacted more than other subpopulations, since SOM+ interneurons have the slowest synaptic kinetics among the considered neuronal types in the model. In addition, it has been estimated that a single GABAergic cell may affect more than a thousand pyramidal cells [48,49], which may explain how the activation of GABAergic neurons may become predominant and exert powerful anti-epiletic effects.

Another prediction of this study is the contributing role of PV+ interneurons stimulation on pre-ictal bursting. More specifically, depolarizing the subpopulation of PV+ interneurons contributes to bursting by increasing the number of spikes during the active phase. Also, as it was discussed above, anodal pulses on PV+ interneurons can prompt the active phase and shorten the frequency of pre-ictal bursts. Such observation is in agreement with a previous study by Assaf and Schiller [50], in which optogenetic activation of PV+ interneurons in the ictal regime had an anti-epileptic effect, but a pro-epileptic effect when they were activated in the inter-ictal regime. More recently, it was discussed that paradoxical effects of PV+ activation shown in [51] could be related to the timing of the neurostimulation [52]. Therefore, our results support that a precise, on-demand (closed-loop) stimulation system is required to deliver stimulation at an optimal timing, and avoid the promotion of epileptiform activity.

Direct brain stimulation for epileptic patients is an ongoing research topic, and unfortunately, the lack of randomized control trials comparing different stimulation protocols hampers obtaining definite results on optimal stimulation protocols [7,53]. Low-frequency electrical and optical stimulation (< 5Hz) applied during interictal phases has been shown to reduce the frequency of interictal spikes and seizure initiation in animal and human studies [41,54]. High-frequency electrical stimulation (>100 Hz) applied during ictal phases has also been shown to terminate seizures [55–57]. Here, we considered the pre-ictal phase, which is between the interictal and ictal phases. We showed that stimulation with a frequency greater than 20 Hz can abort pre-ictal oscillations and keep the system close to background activity by depolarizing the subpopulation of SOM+ interneurons. From our results, the suggested frequency range lies between the ranges of low- and high-frequency stimulations and beyond. This can be due to the fact that the considered epileptogenic phase (pre-ictal) is "in-between" the phases where low-frequency (interictal phase) and high-frequency (ictal phase) stimulations are successful. Furthermore, the spatio-temporal effects of ongoing neural dynamics, synaptic plasticity or differences in the activation functions or neural subpopulations are not included in the model. Nevertheless, our results suggest an alternative stimulation protocol in terms of frequency and timing of stimulation delivery.

It has long been reported that pre-ictal spiking/bursting is an emerging feature of the interictal to ictal transition and is specific to epileptogenic regions. From a mathematical viewpoint, both spiking (a single bump followed by a quiescent phase) and bursting (a sequence of spikes (bumps) followed by a quiescent phase) oscillations in a neural context result from the interaction between the slow and fast variables of a multiple time-scale system. While the type of the oscillation depends on the bifurcation structure of the fast subsystem, it is always the slow subsystem that drives the recurrent transitions between the quiescence and active (spiking) phases [30], here the subpopulation of SOM+ interneurons. Since the essence of spiking/bursting is the same in general sense, stimulation protocols mainly affecting slow oscillations during the pre-ictal phase would abort pre-ictal spiking/bursting activity. In other words, the burst-abortion strategy presented in this paper would also be appropriate to abort spiking. Yet, it is essential to identify the neuronal subpopulations of the brain region under consideration, the connections between these neuronal subpopulations and their roles in such slow-fast regimes to optimize the stimulation frequency, since as shown in this manuscript, subpopulations with slower kinetics are more responsive to pulsed stimulations. For instance, a pre-ictal spiking regime mediated by GABA$_B$ interneurons may be aborted by using lower stimulation frequencies than a pre-ictal spiking regime mediated by GABA$_A$ interneurons, since GABA$_B$ interneurons have slower kinetics than GABA$_A$ interneurons. Depending on the neuroanatomy and neurophysiology of a specific brain region (type of subpopulations and connections between them), activation of specific types of interneurons can be achieved *via* the modulation of different neural targets [40,41,43,58–60].

Slow-fast analysis of the mathematical models of neural systems with complex oscillatory patterns has contributed to discover the roles of biological ingredients [61–70], unveil the fine structures (e.g. excitability thresholds, spike adding mechanisms and subthreshold oscillations …etc.) [19,71–79], and design controllers [80,81]. Recently, response types of brief electrical pulses in coupled neural mass models have been investigated using some elements of slow-fast analysis [82]. In [83] a regime of canard solutions has been reported in sleep/wake transitions in a NMM, also in an extended NMM formulation in [84]. As opposed to [82-84], here we reformulated a widely studied NMM in an explicit slow-fast form and unveiled its three-time-scale structure. During our investigations we also observed canard solutions organizing the transition from slow-wave ($\approx 6$ Hz) to bursting oscillations through a spike-adding mechanism in between. We did not further explore this interesting mechanism since the main purpose of this paper was to understand the perturbation effect on pre-ictal bursting solutions, which are away from the canard regime in the parameter space. Further analysis concerning the classification of slow dynamics near the fold points, canard solutions and spike-adding mechanisms in the line of [20,21,84] are among the possible extensions of this work.

Another possible avenue to extend this work would be to consider the possibility to perform patient-specific bifurcation analyses of epileptiform patterns to propose patient-specific stimulation parameters (most critically, stimulation frequency) that would result in the abortion of the said epileptiform patterns. Current direct brain stimulation protocols in epilepsy use indeed relatively generic parameters, without consideration for the type, localization or extent of the epileptogenic network; a possible factor to explain the lack of consistency for this therapy so far for drug-refractory epilepsy. For our methods to be applicable an adaptive closed-loop detection system, such as a brain responsive neurostimulation system (NeuroPace[TM], NeuroPace, Mountain View, CA, U.S.A.), can be taken into account.

Finally, we should emphasize that the NMM considered here was initially proposed a model for hippocampal activity. As shown in this study, this NMM can reproduce complex oscillatory patterns at the macroscopic level resulting from interaction of district subpopulations with different kinetics. More recently, the model was shown to have a more general scope as the embedded circuitry is valid that of most of the regions at macroscopic level (see [86] and references there in). Thus, appropriately formulated NMMs and the tools presented here can be used to study the complex dynamics observed in other cortical areas and to investigate effects of external perturbations.

## Supporting information captions

**S1 Fig. Periodic orbits of** (4) **and** (8) **for ε=0.01**. The system is put in the bursting regime by taking B=18 and the other parameters are as given in Table 1. (a) Solution of (3) for ε=0.01 for projected on the bifurcation diagram (black curve) of (4) for ε=0 where $v_2$ is treated as a parameter. Stable and unstable solutions are indicated with bold and dashed lines, respectively. The equilibrium points along the black Z-shaped curve are unstable on the middle branch of the curve, between the limit points (LP) $LP_1$ and $LP_2$ (black dots), and on the upper branch between the supercritical Hopf (H) bifurcation points $H_1$ and $H_2$ (green dots). The amplitude of the stable limit cycles is bounded by the green continuous curves connecting the $H_1$ and $H_2$ points in the ε=0 limit. Arrows show the direction of the flow. (b) Solution of (7) for ε=0.01 projected on the bifurcation diagram of (8) (red curve) where $v_2$ is treated as a parameter. Stable and unstable solutions are indicated with bold and dashed lines, respectively. The equilibrium points along the black Z-shaped curve are unstable on the middle branch of the curve, between the $LP_1$ and $LP_2$ limit points (red dots). Arrows show the direction of the flow.

**S2 Fig. Response of a NM block to biphasic balances pulses at 10 Hz.** A NM block reads

$$\ddot{y} = M/\tau \, S\big(I_{ext}(t)\big) - 2/\tau \dot{y}_2 - 1/\tau^2 y_2.$$ Although the ratio of $M/\tau$=1 across the trials, the amplitude of the response is different due to the difference between the synaptic kinetics. The ratio of $M/\tau$ is kept constant. (a) Pulse width is 0.5 ms, pulse amplitude is 1 (arb. unit). (b) Pulse width is 50 ms, pulse amplitude is 0.01 (arb. unit). Increasing the pulse with introduces oscillations around the base line.

Fig. 1

(a)

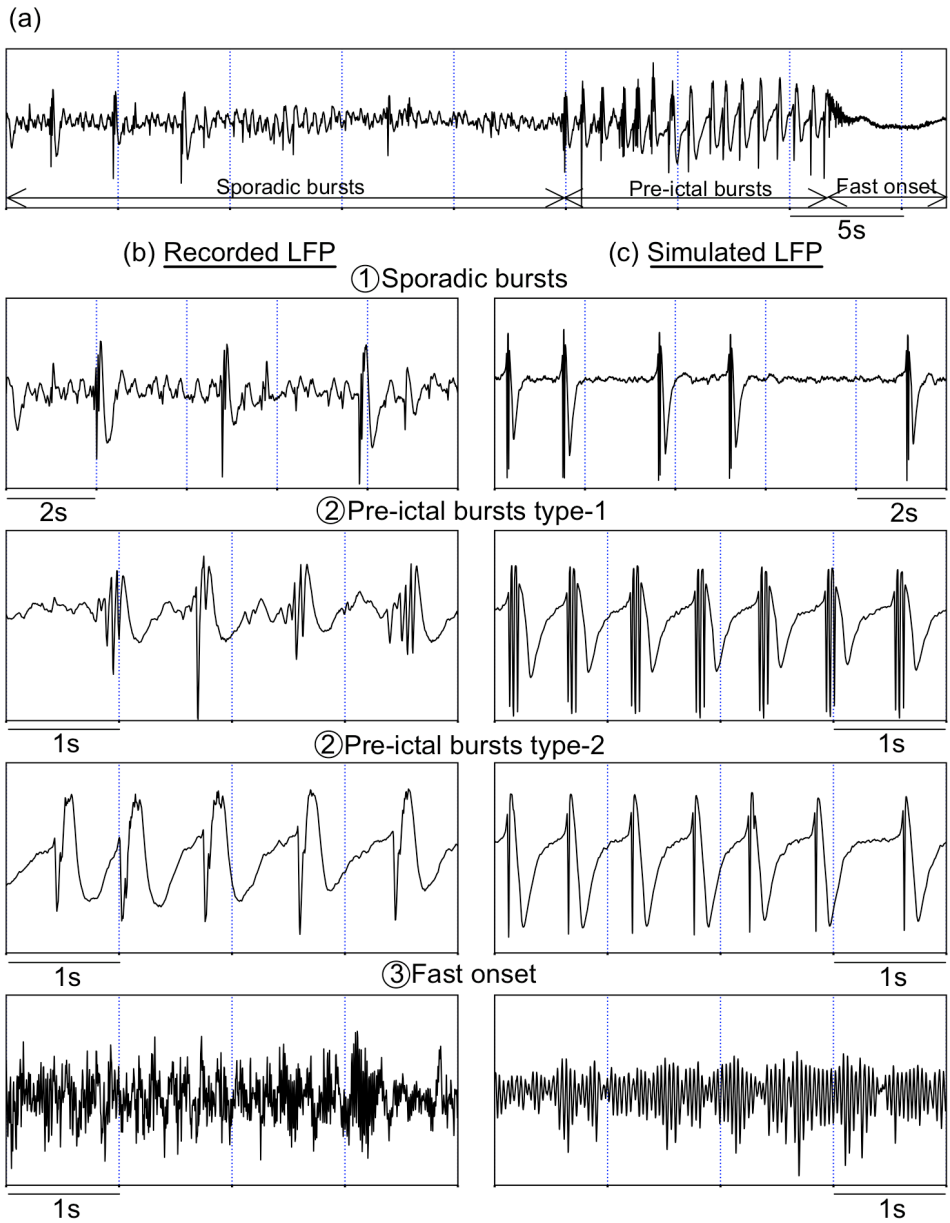

Sporadic bursts | Pre-ictal bursts | Fast onset

5s

(b) Recorded LFP　　　　　　　(c) Simulated LFP

①Sporadic bursts

2s　　　　　　　　　　　　　　　2s

②Pre-ictal bursts type-1

1s　　　　　　　　　　　　　　　1s

②Pre-ictal bursts type-2

1s　　　　　　　　　　　　　　　1s

③Fast onset

1s　　　　　　　　　　　　　　　1s

Fig. 2

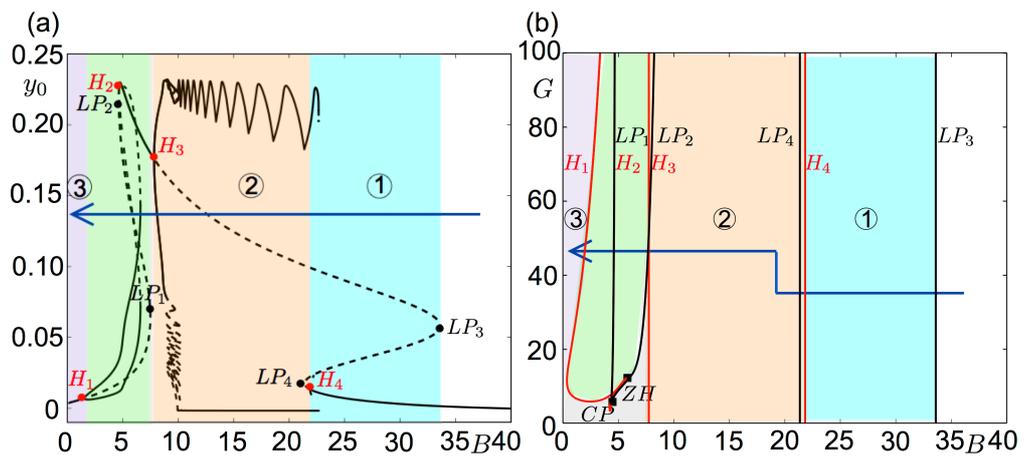

Fig. 3

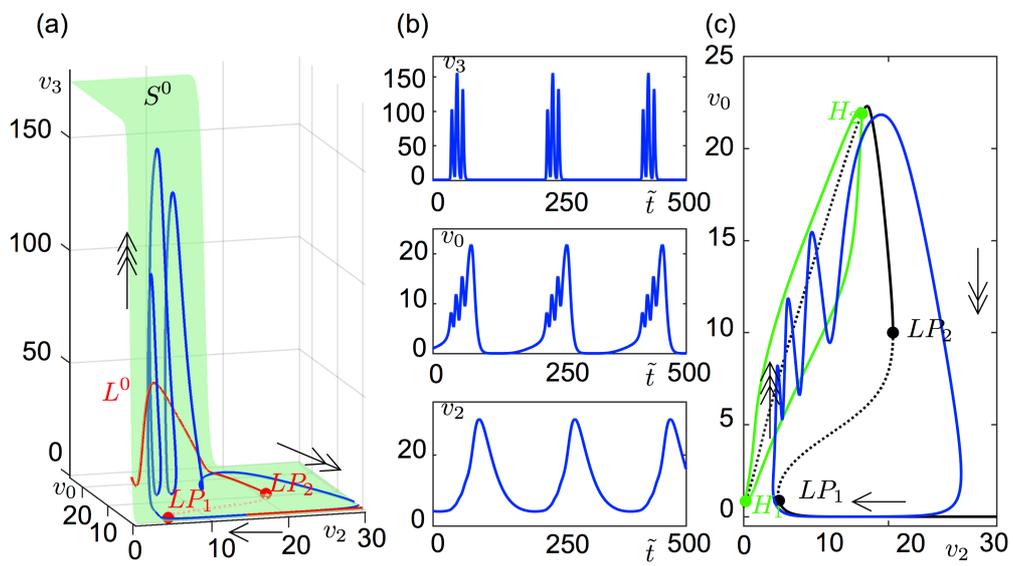

(a) $S^0$, $v_3$, $v_0$, $L^0$, $LP_1$, $LP_2$, $v_2$

(b) $v_3$, $v_0$, $v_2$, $\tilde{t}$

(c) $v_0$, $H$, $LP_2$, $LP_1$, $H$, $v_2$

Fig. 4

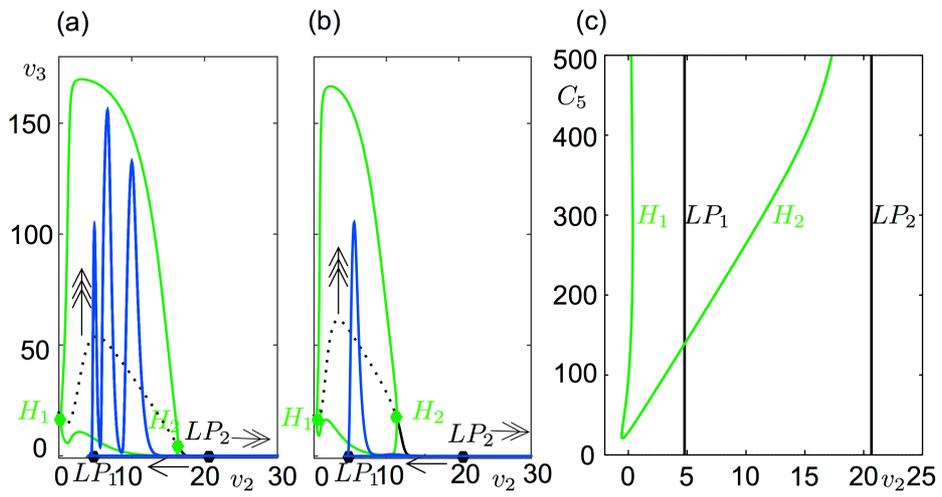

Fig. 5

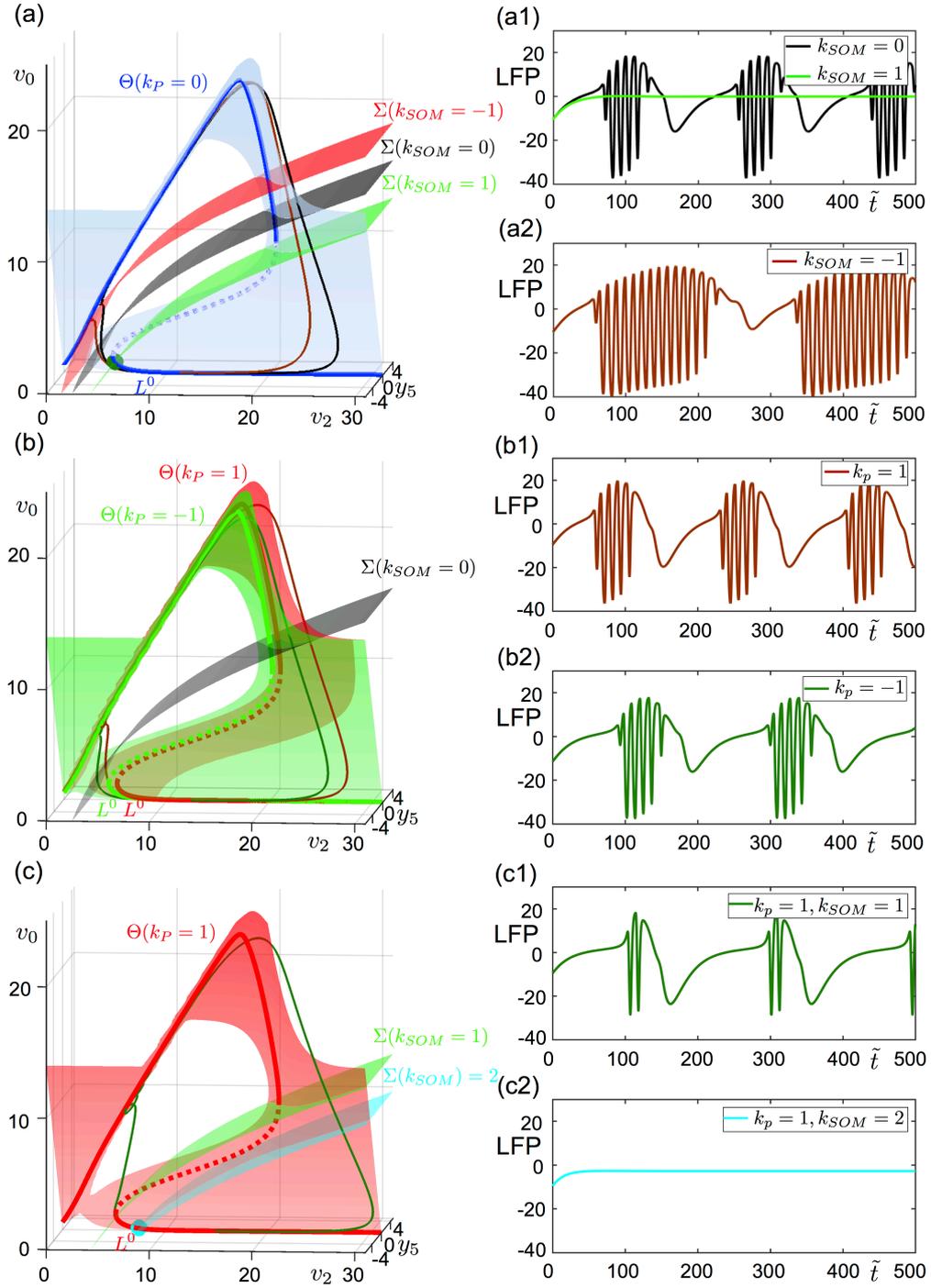

Fig. 6

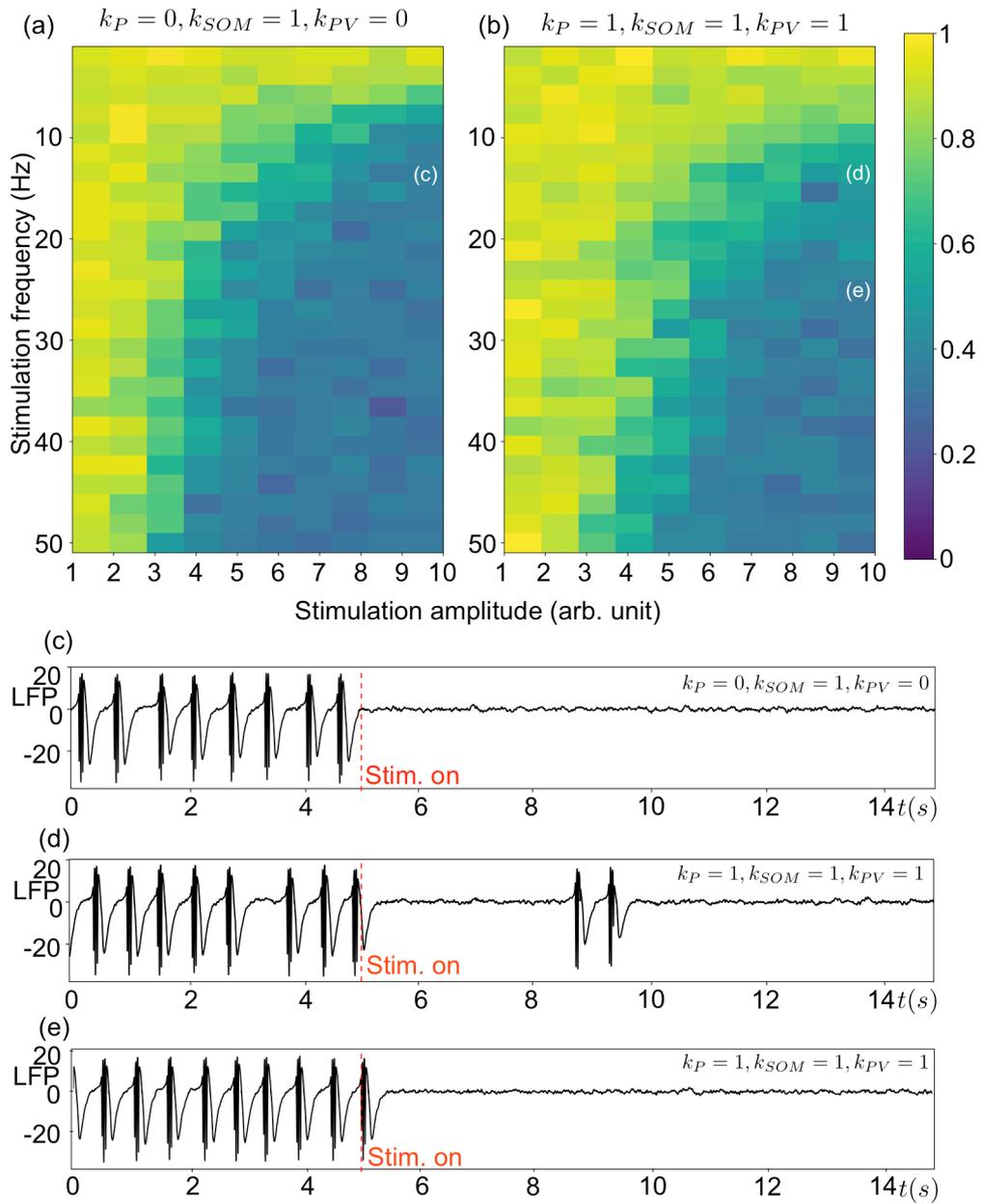

(a) $k_P = 0, k_{SOM} = 1, k_{PV} = 0$
(b) $k_P = 1, k_{SOM} = 1, k_{PV} = 1$

Stimulation frequency (Hz)

Stimulation amplitude (arb. unit)

(c) $k_P = 0, k_{SOM} = 1, k_{PV} = 0$

(d) $k_P = 1, k_{SOM} = 1, k_{PV} = 1$

(e) $k_P = 1, k_{SOM} = 1, k_{PV} = 1$

LFP

Stim. on

$t(s)$

S1fig

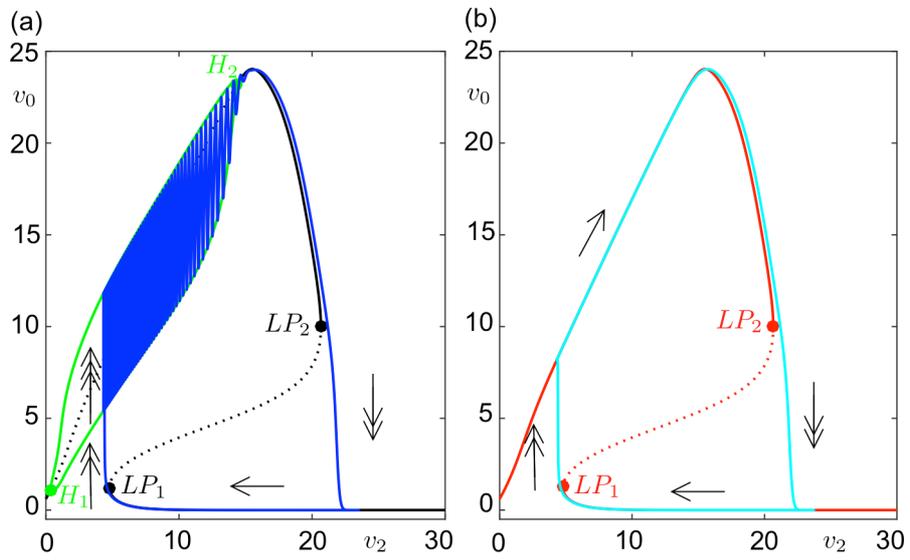

S2fig

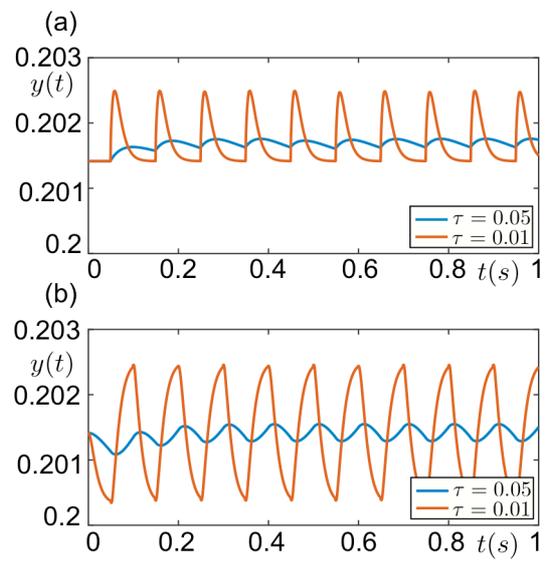